\documentclass[10pt,
twocolumn,
longbibliography,
superscriptaddress,
nofootinbib,
amsmath,amssymb,
aps,
floatfix
]{revtex4-2}

\usepackage[dvipsnames]{xcolor}
\usepackage{hyperref}
\usepackage[utf8]{inputenc}
\usepackage{graphicx}
\usepackage{dsfont}
\usepackage{mathrsfs}
\usepackage{enumitem}
\setlist[itemize]{leftmargin=*}
\setlist[enumerate]{leftmargin=*}
\setlist[enumerate]{itemsep=0mm}

\usepackage{tikz}
\usetikzlibrary{fadings,decorations.pathreplacing}
\tikzset{
    partial ellipse/.style args={#1:#2:#3}{
        insert path={+ (#1:#3) arc (#1:#2:#3)}
    }
}

\usepackage[acronym]{glossaries}
\glsdisablehyper
\makeglossaries

\newacronym{spam}{SPAM}{State Preparation and Measurement}
\newacronym{rb}{RB}{Randomized Benchmarking}
\newacronym{svd}{SVD}{Singular Value Decomposition}
\newacronym{qpt}{QPT}{Quantum Process Tomography}
\newacronym{gst}{GST}{Gate Set Tomography}
\newacronym{asf}{ASF}{Average Sequence Fidelity}
\newacronym{cp}{CP}{Completely Positive}
\newacronym{cptp}{CPTP}{Completely Positive Trace Preserving}
\newacronym{povm}{POVM}{Positive Operator Valued Measurement}
\newacronym{mpo}{MPO}{Matrix Product Operator}

\usepackage{bbm}

\usepackage{mathtools}

\newcommand*{\f}{\frac}
\newcommand*{\mc}{\mathcal}
\newcommand*{\dg}{\dagger}

\newcommand*{\mbb}{\mathds}
\newcommand*{\mscr}{\mathscr}
\newcommand*{\transp}{\mathrm{T}}

\newcommand*{\asfexp}{\mathcal{F}^{(\mathrm{exp})}}
\newcommand*{\dimE}{d_\mathsf{E}}
\newcommand*{\dimS}{d_\mathsf{S}}
\newcommand*{\deff}{d_{\mathsf{eff}}}
\newcommand*{\env}{\mathsf{E}}
\newcommand*{\syst}{\mathsf{S}}
\DeclareMathOperator{\tr}{tr}

\DeclareMathOperator*{\Motimes}{\text{\raisebox{0.25ex}{\scalebox{0.6}{$\bigotimes$}}}}

\DeclareMathOperator*{\Mcirc}{\text{\raisebox{0.25ex}{\scalebox{0.7}{$\bigcirc$}}}}

\usepackage{accents}

\definecolor{C1}{HTML}{2EC4B6}
\definecolor{C2}{HTML}{F46036}
\definecolor{C3}{HTML}{279AF1}
\definecolor{C4}{HTML}{CC4BC2}

\hypersetup{colorlinks=true, linkcolor=C2, citecolor=C2, urlcolor=C2}

\begin{document}
\title{Machine Learning of Average Non-Markovianity from Randomized Benchmarking}

\author{Shih-Xian Yang}
\email{hans.sx.yang@foxconn.com}
\affiliation{Hon Hai Quantum Computing Research Center, Taipei, Taiwan}

\author{Pedro Figueroa--Romero}
\email{pedrofigueroarom@gmail.com}
\affiliation{Hon Hai Quantum Computing Research Center, Taipei, Taiwan}

\author{Min-Hsiu Hsieh}
\email{min-hsiu.hsieh@foxconn.com}
\affiliation{Hon Hai Quantum Computing Research Center, Taipei, Taiwan}

\begin{abstract}
    The presence of correlations in noisy quantum circuits will be an inevitable side effect as quantum devices continue to grow in size and depth. \gls{rb} is arguably the simplest method to initially assess the overall performance of a quantum device, as well as to pinpoint the presence of temporal-correlations, so-called non-Markovianity; however, when such presence is detected, it hitherto remains a challenge to operationally quantify its features. Here, we demonstrate a method exploiting the power of machine learning with matrix product operators to deduce the minimal average non-Markovianity displayed by the data of a \gls{rb} experiment, arguing that this can be achieved for any suitable gate set, as well as tailored for most specific-purpose \gls{rb} techniques.
\end{abstract}

\date{\today}
\maketitle
\glsresetall

%\section{Introduction}
The \gls{rb} protocol is arguably the most dominant and universal paradigm to initially assess the performance of a quantum device. \gls{rb} estimates the average error rates of quantum information processors with a minimal resource overhead and in a manner that is, in principle, insensitive to \gls{spam} errors~\cite{Emerson_2005,Levi_2007,Knill2008}. In particular, the experimental resources to perform \gls{rb} scale only polynomially with the number of qubits being characterized~\cite{PhysRevLett.106.180504, PhysRevA.85.042311}. While limited in scope, it can be considered together with more comprehensive techniques, such as \gls{gst}~\cite{greenbaum2015introduction, Nielsen2021gatesettomography}, which is able to fully characterize the gates involved in a quantum computation, with the trade-off being an exponential scaling of resources needed in system size.

There is currently, however, a regime that is only now beginning to be explored, given the quickly increasing capabilities of quantum devices, namely, so-called non-Markovianity. The term \emph{Markovian} is generally understood as an independence of the state of a system at some given time from its previous outcomes. Classically, this means that for a discrete stochastic process $\{X_t\}$, we have
\begin{equation}
    \mbb{P}(x_{k+1}|x_k,\ldots,x_0) = \mbb{P}(x_{k+1}|x_k,\ldots,x_{k-\ell}),
    \label{eq: Markov classical}
\end{equation}
with $\ell=0$, i.e., the probability to obtain some outcome $x_{k+1}$ at timestep $k+1$ only depends on the outcome $x_k$ at timestep $k$. Whenever this condition in Eq.~\eqref{eq: Markov classical} is not satisfied, we say the process is \emph{non-Markovian} with Markov order $\ell$. The quantum generalization of the concept of Markovianity, however, is not straightforward: probabilities are obtained from quantum states and observations are now inherently invasive, leading to a myriad of conceptual troubles~\cite{milz2020quantum}.

In the context of quantum computation, Markovianity at the gate level is an approximation introduced as a temporal locality of the physical implementation of the gates, assuming that the noise takes the form of a \gls{cptp} map associated, or attached, to each individual gate~\cite{Wallman_2014, helsen2020general}. Nevertheless, we know from the theory of open quantum systems that non-Markovianity is the norm and Markovianity the exception~\cite{vanKampen1998}; even when it can typically be assumed~\cite{2019almostmarkovian}, the Markovianity approximation for the noise inevitable becomes invalid for quantum circuits incorporating a large amount of qubits or long sequences of gates, with most of the current techniques quickly failing and becoming unreliable~\cite{osti_1671379, Proctor2020}.

The standard \gls{rb} protocol, detailed in Appendix~\ref{Appendix: rb protocol}, proceeds by implementing a sequence of noisy random quantum gates on a given initial state, then applying the compiled ideal inverse sequence, and finally analyzing the average of the expectation value of a given measurement element. A particularly relevant case is that when the gates belong to the multi-qubit Clifford group~\cite{meier2018randomized}, as these have well-known analytical properties and can be simulated efficiently~\cite{gottesman1998heisenberg, gottesman1997stabilizer, PhysRevA.70.052328}. More generally, as long as the gates belong to a unitary 2-design\footnote{ A unitary $t$-design is a set of unitaries reproducing up to $t$ moments of the unitary group with the uniform (Haar) measure. The multi-qubit Clifford group is one such example~\cite{webb2016,graydon2021clifford} and the one typically used in standard \gls{rb}.}, and assuming Markovianity, gate-independence, time-independence, and trace-preservation properties on the noise, the data produced by \gls{rb} follows an exponential decay whereby the error rates are encoded in the rate of such decay. This means that, in principle, all such \gls{rb} experiments have to do is fit an exponential to the outputs in order to extract average gate fidelities.

While \gls{rb} remains tractable upon relaxing most assumptions~\cite{helsen2020general}, the outputs of \gls{rb} in the non-Markovian case become a non-trivial function of the space mediating the temporal correlations~\cite{figueroaromero2021randomised, figueroaromero2022general}, as we briefly summarize in Appendix~\ref{Appendix: rb clifford nM}, rendering deviations from the standard exponential decay. In particular, quantifying the amount of average non-Markovianity, or other memory features giving rise to this non-exponential behavior, remains quite challenging in practice.

In this manuscript, we connect the machine learning technique for tensor networks developed in~\cite{stoudenmire2016} to the process tensor framework for non-Markovian processes~\cite{PhysRevA.97.012127}, allowing us to extract information such as the average amount of non-Markovianity and noise memory-length from any given \gls{rb} experiment's data alone. Other than assuming gate-independence in the noise, our method can be employed with any multi-qubit gate set admitted by standard \gls{rb} and is able to diagnose \gls{spam} correlations. Moreover, our approach can easily be tailored for specific-purpose \gls{rb}~\cite{helsen2020general} and accommodate for extensions beyond noise-intermediate scales. Our method serves as a quick diagnose of average non-Markovianity, which can be further complemented by more computationally expensive techniques, such as Non-Markovian Process Tensor Tomography~\cite{PRXQuantum.3.020344}.

\section{RB for non-Markovian Noise as a contraction of MPOs}
\begin{figure}[t!]
    \centering
    \includegraphics[width=\linewidth]{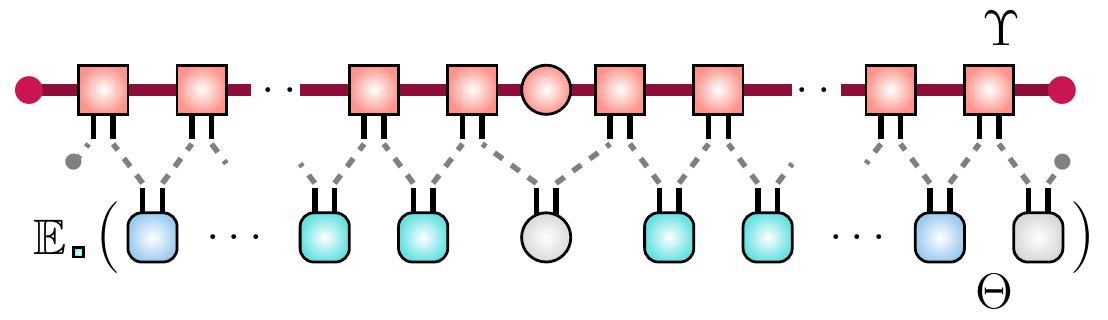}
    \caption{\textbf{Randomized Benchmarking as a contraction of Matrix Product Operators.} On top, the noise tensor $\Upsilon$, with unitary noise maps $\Lambda_i$ (red squares) and the fiducial state $\rho_\env$ (red circle), is correlated via an environment of bond dimension (red thick line) at most equal to $\dimE$, while below a control tensor $\Theta$ contains all the gates $\mc{G}_i$ (green/blue rounded squares), the initial state $\rho_\syst$ (gray circle) and the measurement $M$ (gray rounded square). Averaging over gates, denoted $\mbb{E}$ is done over $\Theta$ (resulting effectively in a sum of \gls{mpo}s). In dashed lines is denoted the contraction order, while bullets denote contractions at the edges.}
    \label{fig: upsilon omega}
\end{figure}
We may generally describe non-Markovianity as mediated by an external environment: in the quantum case, this corresponds to a Hilbert space $\mscr{H}_\env$ part of a total composite $\mscr{H}_\env\otimes\mscr{H}_\syst$, with the labels $\env$ and $\syst$ corresponding to environment and system of interest, respectively. A non-Markovian \gls{rb} sequence may be described as follows. An initial state $\rho_\syst$ is prepared on $\syst$, which then can interact with $\env$, initially in some fiducial state $\rho_\env$, via a \gls{cp} map $\Lambda_0$ acting jointly on $\syst\env$, until a gate $\mc{G}_1$ is applied on system $\syst$, followed by a \gls{cp} map $\Lambda_1$, and so on, until the undo gate $\mc{G}_{m+1}=\mc{G}_1^{-1}\circ\cdots\circ\mc{G}_m^{-1}$ is applied followed by noise $\Lambda_{m+1}$ and a \gls{povm} element $M$ solely on $\syst$. Sequences such as this one, together with all statistics and $(m+1)$-point correlations, are described by the process tensor framework~\cite{PhysRevA.97.012127, PhysRevLett.120.040405, Milz2020kolmogorovextension, taranto2019memory, PhysRevLett.123.040401, milz2020quantum, White_2020, nurdin2021heisenberg, white2021nonmarkovian}, whereby the inherent noise dynamics of the full $\syst\env$ can be given in a tensor $\mscr{Y}_{m+1}$, and the sequence of gates in a tensor $\mscr{G}_{m+1}$.

The expectation of $M$, averaged over the quantum gates, is known as the \gls{asf}, and can now be computed as
\begin{equation}
    \mc{F}_m = \tr\left\{M\tr_{\overline{\syst}}\left[\mscr{Y}_{m+1}\mbb{E}\left(\mscr{G}_{m+1}^{\mathrm{T}}\right)\right]\right\},
    \label{eq: asf pt general}
\end{equation}
where $\mbb{E}$ is an average over uniformly sampled gates $\mc{G}_i$, $\tr_{\overline{\syst}}$ is partial trace over all auxiliary systems in the tensors, except $\syst$, and the corresponding tensors can be defined through their Choi state representation~\footnote{ The Choi state representation~\cite{watrous2018theory} is defined for a quantum channel $\Phi$ as $\Upsilon_\Phi:=(\mc{I}\otimes\Phi)\psi$, where $\mc{I}$ is an identity channel, i.e., by letting the channel act on half a maximally entangled state. This is generalized in the process tensor as in Eq.~\eqref{eq: PT choi state}, with Eq.~\eqref{eq: gates choi state} being a particular uncorrelated case.} by
\begin{align}
    \mscr{Y}_{m+1} &:= \tr_\env\left[\Mcirc_{i=1}^{m+1}\left(\Lambda_{i}\circ\mscr{S}_{i}\right)\left(\rho\otimes\psi^{\otimes{m+1}}\right)\right] \label{eq: PT choi state}\\
    \mscr{G}_{m+1} &= \mbb1_\syst\otimes\Motimes_{i=1}^{m+1}\left(\mbb1_{\mathsf{A}_{i}}\otimes\mc{G}_{i}\right)\psi^{\otimes{m+1}} \label{eq: gates choi state},
\end{align}
where here $\circ$ is a composition of maps, $\rho:=\Lambda_0(\rho_\syst\otimes\rho_\env)$ and $\psi=\sum|ii\rangle\!\langle{jj}|$ are unnormalized maximally entangled states defined in auxiliary spaces $\mathsf{A}_1\mathsf{B}_1\cdots\mathsf{A}_{m+1}\mathsf{B}_{m+1}$; the map $\mscr{S}_i$ is a swap between system $\syst$ and the $i$\textsuperscript{th} ancillary space $\mathsf{A}_i$.

The \gls{asf} is the analytical form of the data estimated by the \gls{rb} protocol. Whenever the underlying noise is approximated as Markovian, time-independent, gate-independent and trace-preserving, it can be seen that the \gls{asf} in Eq.~\eqref{eq: asf pt general} reduces to $\mc{F}_m=Ap^m+B$, where $p$ is directly related to the average gate fidelity of $\Lambda$ with respect to the identity, and $A$ and $B$ are constants only depending on~\gls{spam} errors.

The Choi state representation of the process tensor is simply a many-body state (up to normalization) translating its temporal correlations into spatial ones; naturally, as any other many-body operator, it admits a \gls{mpo} decomposition~\cite{PhysRevA.97.012127, milz2020quantum}, i.e., it can be expressed as a chain of contractions of single operators. We can thus rearrange the noise process, i.e., all $\Lambda_i$ \gls{cp} maps and the fiducial state $\rho_\env$, as a \gls{mpo} $\Upsilon_{m+1}$, and the control process, i.e., the gates $\mc{G}_i$, the initial state $\rho_\syst$, and the measurement element $M$, respectively, as another \gls{mpo}, $\Theta_{m+1}$, such that
\begin{align}
    \mc{F}_m &= \tr\left[\Upsilon_{m+1}\Theta_{m+1}^\dg\right].
    \label{eq: asf main MPOs}
\end{align}

This is depicted graphically for the case of unitary noise in Fig.~\ref{fig: upsilon omega}, where the free legs in each tensor represent the corresponding degrees of freedom in $\syst$ and the thick line joining each node of $\Upsilon$ represents its so-called bond dimension. We derive both \gls{mpo}s explicitly in Appendix~\ref{Appendix: tensors}, with which the inner product in Eq.~\eqref{eq: asf main MPOs} can be written explicitly as a sum over their degrees of freedom in $\syst$.

The \gls{mpo} decomposition of the process tensor opens up new possibilities to efficiently study features of non-Markovian processes~\cite{PhysRevLett.122.160401}. While the dimension of $\mscr{Y}_{m+1}$ in Eq.~\eqref{eq: PT choi state} grows exponentially in sequence length, the construction of the \gls{mpo} $\Upsilon_{m+1}$ can be made with a complexity scaling linearly in sequence length, and similarly for the control tensor $\Theta_{m+1}$. Furthermore, while generically the dimension $\dimE$ of the environment is large, the bond dimension of $\Upsilon_{m+1}$ will usually be much smaller, as often only a finite part of the environment interacts with the system at a given time~\cite{luchnikov2018dimension, PhysRevLett.122.160401}. The bond dimension of $\Upsilon_{m+1}$ corresponds precisely to this minimal dimension, $\deff$, for an environment to propagate memory through different timesteps. In the context of \gls{rb}, then, deviations from an exponential decay of the \gls{asf} can be directly tied to the bond dimension of the noise tensor.

We now show how these features can be extracted from a \gls{rb} experiment's data alone, thereby quantifying average non-Markovianity rates, by employing the machine learning techniques devised in~\cite{stoudenmire2016}.

\section{Supervised Learning for non-Markovian RB}
The use of \gls{mpo}s has already proven useful for the numerical and conceptual investigation of open quantum systems, e.g., in~\cite{PhysRevLett.105.050404, Lovett2018, PhysRevA.94.053637,PhysRevLett.123.240602, PhysRevLett.122.160401}, and are still an active subject of investigation. Tensor networks, of which \gls{mpo}s are particular cases, naturally find applications for machine learning tasks~\cite{stoudenmire2016,PhysRevX.8.031012,Klus_2019,Liu_2019, Araz_2021, PhysRevB.103.125117,Wang21, PhysRevResearch.3.023010}, and can further serve to learn full process tensors~\cite{PhysRevA.102.062414}. Here we show a way to train a model that allows to learn the effective average non-Markovianity from a \gls{rb} experiment's data.

The outputs of a \gls{rb} experiment are estimations of the \gls{asf} in Eq.~\eqref{eq: asf main MPOs}. Let $\{\asfexp_n\}_{n=1}^m$ be such outputs, which are real numbers between 0 and 1 with a given uncertainty, for an experiment with a given gate set $\{\mc{G}_i\}_{i=1}^m$, target initial state $\rho_\syst$ and measurement $M$. Thus we aim to be able to estimate a model for the underlying noise giving rise to such statistics: in particular we want to extract an effective environment dimension and an effective noise memory length. To do so, we must optimize a cost function relating the known outputs with our model; concretely, we choose to optimize the quadratic cost
\begin{align}
    C &:= \frac{1}{2}\sum_{n=1}^m\left(\mc{F}_n - \asfexp_n\right)^2,
    \label{eq: cost function}
\end{align}
by adapting the sweeping algorithm of~\cite{stoudenmire2016} for the \gls{mpo} representation of the process tensor as follows.

The sweeping algorithm is an iterative method that proceeds by updating pairs of nodes of the corresponding \gls{mpo} model by gradient descent.
We adapt this method to the process tensor by fixing an environment dimension and demanding physicality of the noise, i.e. that the nodes in the noise tensor $\Upsilon$ constitute at least \gls{cp} maps. We choose to ensure this by assuming a closed $\syst\env$ system, i.e., that all the nodes $\lambda_i$ in $\Upsilon$ are unitary~\footnote{ Alternatively, one may demand the noise to be \gls{cp}, i.e., for the nodes to be a sum of Kraus operators. While generally this may relax how large the initial dimension of $\env$ must be, the trade-off is needing to fit, in principle, all $d_\env^2{d}_\syst^2$ Kraus operators.}.
Demanding the learned \gls{mpo} to be physical is the major difference between our method and the original sweeping algorithm.

Projected gradient descent is  a method to solve a constraint optimization problem: it projects the gradient onto the constraint set.
In our method, we project a square matrix $X$ onto an unitary constraint set, $Q_U$.
This projection $\pi(X)$ is implemented by an optimization $\pi(X)=\arg \min _{Q \in Q_U}\|X-Q\|^{2}$.
Practically, this projection has a simpler implementation as shown in~\cite{984753}:
the solution can be reached by taking \gls{svd} on the matrix $X$, resulting in two unitaries $U, V^\dg$ and a singular matrix $S$.
The solution is simply given by replacing the singular matrix by an identity one, $\pi(X) = UV^\dg$.
% According to~\cite{984753}, a matrix $X$ being projected into a unitary constraint set $Q$, there exists a corresponding unique matrix, $UV^\dg$\com{?}, where $U$ and $V^\dg$ are the unitaries obtained from the \gls{svd} on $X$\rem{ and $S$ is the singular value matrix}.}

% \sxre{The sweeping algorithm is a powerful algorithm designed to capture the untold feature of data.
% However, our task has some known features depending on the assumptions of a noise model.
% For example, a time-independent noise model implies all the noise $\lambda_i$ in $\Upsilon$ are the same.
% Our method could be tailored for specific purpose routines by wisely incorporating this additional knowledge.}

Briefly summarized, our algorithm proceeds as follows (full detail can be seen in Appendix~\ref{Appendix: RB sweeping algorithm}):
\begin{enumerate}
    \item Contract a neighboring pair of node tensors $\lambda_i$, $\lambda_{i+1}$ in $\Upsilon$ to form a single joint node, $L_{i,i+1}$.
    \item Compute the gradient of the cost function with respect to the joint node, $\Delta{L}_{i,i+1}$, and update the joint node as $L^\prime_{i,i+1} = L_{i,i+1}+\alpha\Delta{L}_{i,i+1}$ with some learning rate $\alpha$.
    Methods such as Adagrad~\cite{JMLR:v12:duchi11a} or Adam~\cite{kingma2014adam} allow for an optimal choice of $\alpha$.
    \item Split the updated joint node by a higher dimensional \gls{svd}~\cite{Bridgeman_2017} and decompose a tensor to be the multiplication of  as $L^\prime_{i,i+1}=(A\sqrt{S})(\sqrt{S}B^\dg)$, where $(A\sqrt{S})$ and $(\sqrt{S}B^\dg)$ are both $\dimE\dimS^2$ square matrices.
    Keeping the columns corresponding to the largest $\dimE$ singular values, truncate the unitaries $A\sqrt{S}\to{\tilde{a}}$, $\sqrt{S}B\to{\tilde{b}}$ to $(\dimE\dimS^2\times\dimE)$ matrices.
    And reshape the truncated matrices to square $\dimE\dimS$ matrices, $\tilde{a}\to{a}$, $\tilde{b}\to{b}$.

    \item Project the truncated matrices to be unitary.
    We follow~\cite{984753} by performing a further \gls{svd} as $a=U_aS_aV_a^\dg$ and $b=U_bS_bV_b^\dg$.
    The projections of $a$ and $b$ are $U_aV_a^\dg$ and $U_bV_b^\dg$, respectively. Multipy them to form a new node $\lambda\to{U}_aV_a^\dg{V}_b{U}_b^\dg$.
    
    \item Update \emph{all} the nodes in $\Upsilon$ as $\lambda\to{U}_aV_a^\dg{V}_b{U}_b^\dg$ and $\lambda^\dg\to{U}_bV_b^\dg{V}_a{U}_a^\dg$. By updating all the nodes we thereby impose time-independence, which speeds up the update process and avoids over-fitting\footnote{ Whenever this restriction is too strong (i.e., minimization of $C$ is not achieved), this may be compensated by increasing the fiducial environment dimension. See Sec.~\ref{sec: conclusions} for a discussion on this issue.}.
    
    \item Return to point 1 with another pair of nodes, aiming to minimize the cost function $C$.
\end{enumerate}

When the fitting model converges to the experimental data, according to a user-specified criterion, and the cost function in Eq.~\eqref{eq: cost function} is minimized, the structure of the resulting tensor will reflect whether the data is Markovian or not. This shall occur either within its bond-dimension, or equivalently in the form of the nodes themselves (e.g., whether they are correlated with the whole of $\env$ or not). The learned noise model itself will be irrelevant and should be simply thought of as an effective average model giving rise to the experimental \gls{rb} data. Then, the very first step to test this is that employ this method to discriminate a non-Markovian noise model from a Markovian one. An effective, albeit artificial, model of the noise tensor, reflecting the average non-Markovianity of the data, or lack thereof, can then be learned for any choice of initial states, measurements and/or gate sets.

We now show two proof-of-principle examples that demonstrate our method.

\section{Numerical examples}
\label{sec: numerical}
In the following, we start by performing standard \gls{rb} with Clifford gates on two different noise models and estimating the corresponding \gls{asf} over 100 samples. This will serve as our \gls{rb} data, on which we aim our method to estimate the minimal environment size giving rise to it and some effective noise models which need not match the true noise other than precisely in the required environment size.

In both cases, for our learning model, we consider one qubit as our system $\syst$ and one qubit as environment $\env$, initially on state $\rho=|00\rangle\!\langle00|$ and a projective measurement $\mathcal{M}=|0\rangle\!\langle0|$. The same procedure can be performed for distinct gate sets, as well as initial states and measurements to generalize the learning model.

We observed sudden jumps in the cost function in both cases.
This might be a consequence of the time-independence constraint we imposed, suddenly updating it away from the minimum.
A more detailed discussion on this issue is in Appendix~\ref{Appendix: interpret_cost}.
Updating only a pair of nodes in a single iteration could provide a better fit at the expense of an increase in iterations of as many times as the sequence length of the experiment; we discuss this possibility in Sec.~\ref{sec: conclusions}

% \sxre{
% This issue\rem{s} might result from the sweeping. Since we impose \rem{the assumption} that all the nodes $\lambda_i$ in $\Upsilon$ are the same, it seems more reasonable to \com{*}take the gradient of the whole $\Upsilon$\com{* What does this mean?} rather than part of it but replace all the $\lambda_i$\com{?}.
% \rem{This inflexibility under the assumption that all the nodes $\lambda_i$ in $\Upsilon$ are the same is the weakness of our method.}
% \rem{It}\add{This time-independence constraint we impose might} lead\rem{s} to the sudden\rem{ly} jump\add{s in the cost function} and \add{the} update\add{s} away from the minimum.\com{Move this to the beginning of the section, it applies to both examples}
% \rem{But unfortunately,}\add{However,} tak\rem{e}\add{ing} the gradient of the whole $\Upsilon$\com{Again, what does this mean? You mean the gradient of C w.r.t the whole $\Upsilon$?} will los\add{e}\rem{s} the advantage of \add{the} \gls{mpo} decomposition and consume\rem{s} \add{an} exponentially increasing \rem{(with sequence length)} memory \add{in increasing sequence lengths}.}

\subsection{Markovian phase flip noise model}
\label{subsec: learned M}
\begin{figure}
    \centering
    \includegraphics[width=\linewidth]{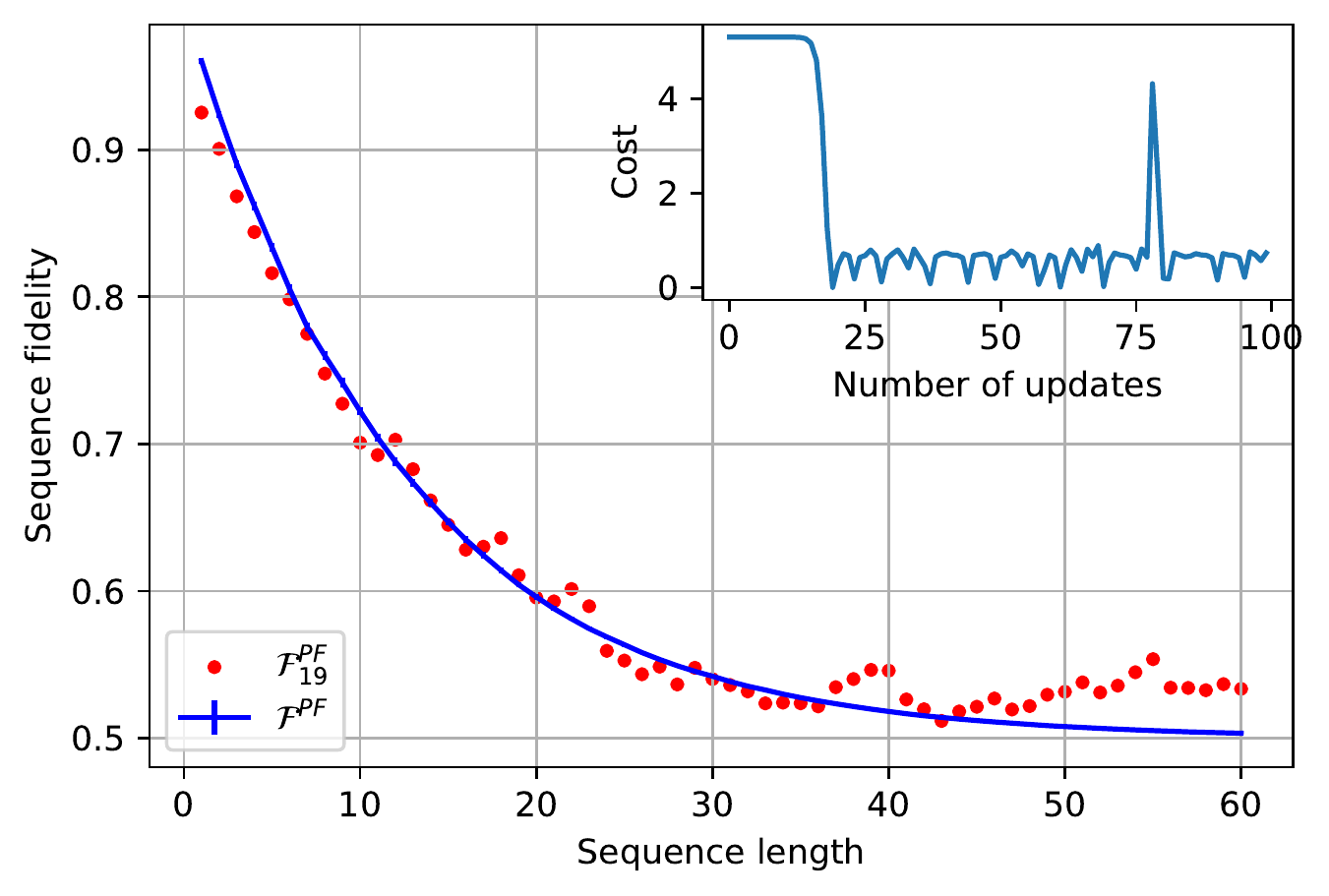}
    \caption{\textbf{Learned \gls{asf} for Markovian phase flip noise.} The blue line is the \gls{asf} from the noise model and the error bar is the standard deviation divided by square root of sample numbers (too small to be seen).
    The red dots are the output of our method, obtained by Adagrad~\cite{JMLR:v12:duchi11a} with initial learning rate $\alpha=10^{-5}$.
    The superscript in $\mathcal{F}^{PF}_{19}$ indicates the noise model and the subscript is the number of iterations to reach this result.
    The subplot shows the change of cost with iterations, the $19^\textit{th}$ iteration gives the \gls{asf} data shown in here.}
    \label{fig:pf_asf}
\end{figure}

% To contrast with We consider purely Markovian \gls{rb} data, 
We consider the single-qubit phase flip channel,
$\mathcal{E}(\cdot) = \sum_{i=0}^1 E_i (\cdot) E_i^{\dg}$, where $E_0 = \sqrt{1-p}
\mbb1$ and $E_1 = \sqrt{p} {Z}$, where $p$ is the phase flip probability, which we fix to $p=0.06$. The corresponding numerical \gls{asf}, the learned points and the cost function as a function of iterations is depicted in Fig.~\ref{fig:pf_asf}. The learned points converged in 19 iterations as the cost function approached a local minimum. The cost function can be seen to first approach such minimum, although afterwards it increases and displays at least one jump far from the minimum.

Crucially, the learned unitary noise has the form of a direct sum, $\lambda = \mbb1_\env\oplus u$ where $u$ is a unitary  solely on $\syst$, implying uncorrelated noise. As shown in Appendix~\ref{Appendix: interpret_M} in another example with an amplitude damping noise model, this is a consistent feature of the method when the noise is Markovian. This reflects that no environment is needed, as there is no temporal correlations, or memory, between time-steps within the noise. This might then be generalized both to identify an \gls{asf}'s data as Markovian or non-Markovian, as well as to more generally infer the minimal size of the correlation needed with an environment to produce such data.

\subsection{Two qubit non-Markovian spin noise model}
\label{subsec: learned nM}
\begin{figure}
    \centering
    \includegraphics[width=\linewidth]{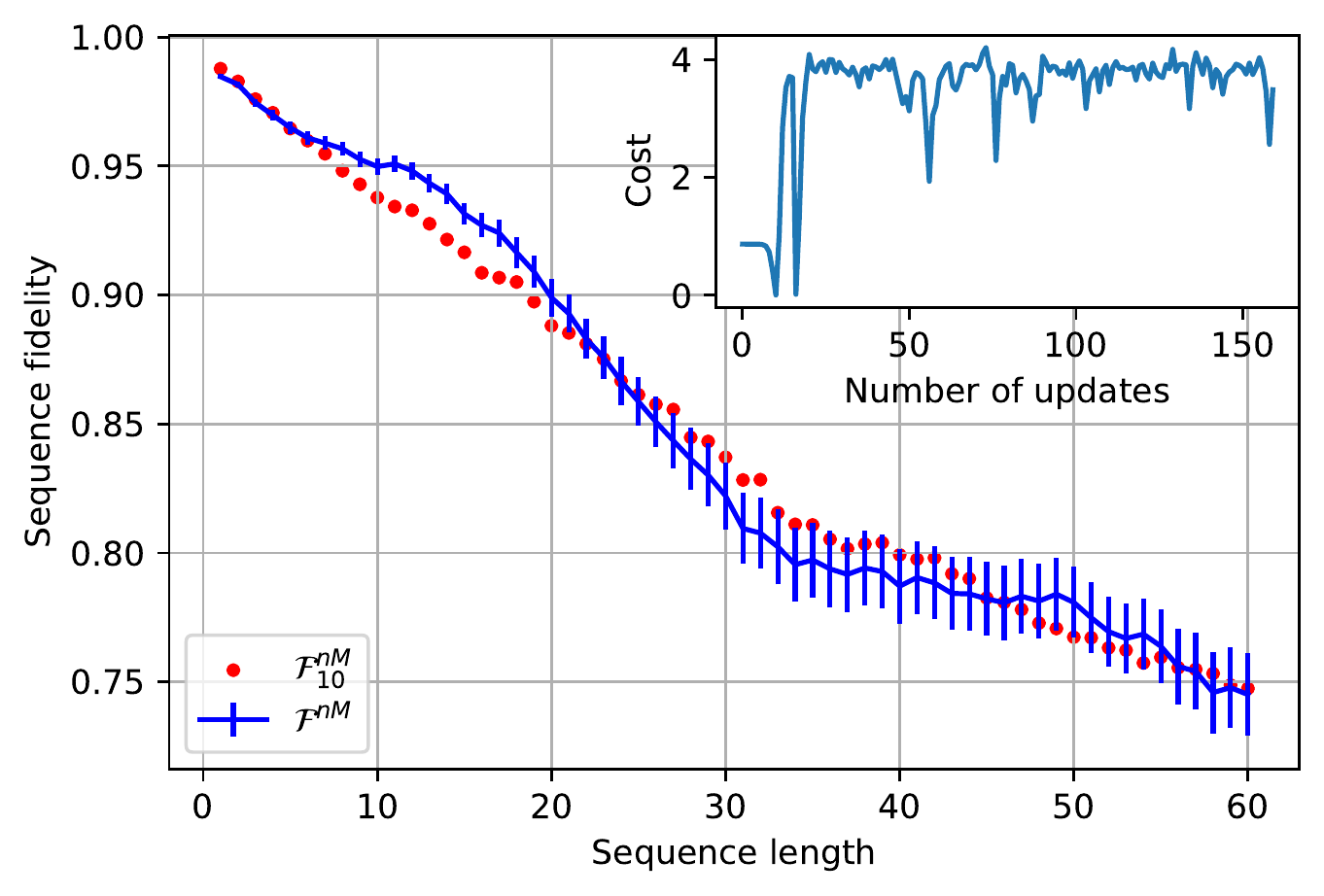}
    \caption{\textbf{Learned \gls{asf} for two qubit fully non-Markovain spin noise model.} The blue line is the \gls{asf} from the non-Markovian noise model and the error bar is the standard deviation divided by square root of sample numbers.
    The red dots are the output of our method, obtained by Adam~\cite{kingma2014adam} with learning rate $\alpha=0.001$ and Adam decay rates $\beta_1=0.9$, $\beta_2=0.99$.
    The superscript in $\mathcal{F}^{nM}_{10}$ indicates the noise model and the subscript is the number of iterations to reach this result.
    The subplot shows the change of cost with iterations, the $10^\textit{th}$ iteration gives the \gls{asf} data shown in here.
    % \sxre{The reasons underlying this fast convergence is that the $\syst\env$ noise is close to identity.}
    % \sxre{--- To make things simple and clear, I moved it here. Although updating all the nodes also contributes, but it is on the level of the algorithm. ---}
    }
    \label{fig:nM_asf}
\end{figure}

We consider the two qubit fully non-Markovian spin noise model from~\cite{figueroaromero2021randomised}, given that~\gls{rb} exponential deviations on it are manifest. It is a static unitary noise model $\Lambda(\cdot) = \lambda(\cdot)\lambda^\dg$, where $\lambda=\exp(-i\delta H)$ for small positive $\delta$. The Hamiltonian is given by the two-spin interaction $H=J X_{1} X_{2}+h_{x}\left(X_{1}+X_{2}\right)+h_{y}\left(Y_{1}+Y_{2}\right)$, with $X_i$ and $Y_i$ being respective Pauli matrices acting on the $i$\textsuperscript{th} site. The \gls{asf} data, $\mathcal{F}^{nM}$ in Fig.~\ref{fig:nM_asf}, is generated by the model with 
$J=1.2, h_{x}=1.17, h_{y}=-1.15$ and $\delta = 0.05$. The cost function updates away from the minimum after two jumps: we discuss this issue further in Appendix~\ref{Appendix: interpret_cost}.

We expect that the learned noise has a form which indicates the interaction among $\syst$ and $\env$.
As shown in Appendix~\ref{Appendix: interpret_nM}, it can be seen that the environment qubit is always needed.
While our method correctly keeps the environment qubit, it can be seen that the deviations from an exponential between the sequence lengths $10$ to $20$ in Fig.~\ref{fig:nM_asf} are not completely resolved by the fit. This might again be related to the time-independence constraint, which we discuss further below.
% As we mentioned on the beginning of this section,

% The method converged in $10$ iterations as shown in Fig.~\ref{fig:nM_asf}.
% The reasons underlying this fast convergence are two-fold: one being that the $\syst\env$ noise is close to identity, and the other being that the nodes are all updated in a single iteration. 
% Updating a pair of nodes in a single iteration would provide a better fit at the expense of an increase in iterations of as many times as the sequence length of the experiment; we discuss this possibility below in Sec.~\ref{sec: conclusions}.}

% This is also indicated in the subplot of the cost in Fig.~\ref{fig:nM_asf}.
% The cost starts small since the $\syst\env$ the noise is close to identity but it later jumps up and no signs to go down.

% This might result from the structure of the way we update.
% Unlike the learned Markovian noise, it could not be decomposed to a form of direct sum.

% The raw outputs for the noise matrices and further detail can be found in Appendix~\ref{Appendix: interpretation}.

\glsresetall
\section{Conclusions and Discussion}\label{sec: conclusions}
We propose a method to deduce the average non-Markovianity, as specified by the minimal environment size, within the underlying noise dynamics producing a given \gls{rb} experiment's data. We demonstrate numerically in two proof-of-principle examples that our method is able to distinguish between the \gls{rb} data of a Markovian and a non-Markovian noise model. We furthermore argue how this method is more generally able to deduce the minimal effective environment dimension required to produce a given \gls{rb} data, as this becomes manifest in the correlation structure of the learned effective noise.

While in this manuscript we present examples restricted to time-independent noise, our method is amenable to be extended for more complex cases such as a finite length non-Markovian noise. This presents new challenges, as the method might naively overfit the data by simply allowing distinct noise $\Lambda_i$ solely on system $\syst$ such that a decay of the form $\mc{F}_m=Ap_1p_2\cdots{p}_m+B$ is obtained (unless the data is purely non-Markovian and displays a fidelity increase in sequence length~\cite{figueroaromero2022general}). Another challenge is that time-dependence would impose a substantial computing overhead, as we would need at least the same amount of iterations of the sweeping algorithm as the given sequence length to do a single update of all the nodes. The constraint to time-independence (described in detail in Appendix~\ref{Appendix: RB sweeping algorithm}), is imposed in the method by replacing \emph{all} the noise nodes by the updated one on each iteration of the sweeping algorithm. Similarly, while we opt to enforce the learning method to be constrained to physical noise by demanding the full $\syst\env$ to be a closed system, this can be relaxed to simply demanding that the noise is represented at least by a \gls{cp} map on $\syst\env$. While this may alleviate having to set a larger dimension of $\env$ in case the real minimal such dimension is larger, it also imposes an overhead in having to fit, in principle, all $\dimE^2\dimS^2$ Kraus operators.

A more fundamental way to improve the foreseen efficiency problem might be doing gradient descent on all the tensors at the same time.
This can be achieved by the direct gradient descent (DGD) technique in~\cite{barratt2022improvements}. They improve the exploding or vanishing gradients for tensor networks via an initialization scheme.
Transforming our task to a DGD problem would be a worth exploring approach.

In this work, we show that it remains  feasible to extract relevant information from the mere simplicity a \gls{rb} experiment's data, namely, about directly inaccessible temporal correlations within a device's quantum noise. Our method can furthermore be extended and tailored for specific purpose routines; this is increasingly relevant as the development of quantum devices continues to scale up, with non-Markovianity remaining one of the major challenges largely unresolved.

% \sxre{For \gls{asf} data from a \gls{rb} protocol with sequence length $m=60$, and assuming one qubit in $\syst$ and one in $\env$. It takes about $7.5$ hours to complete $100$ iterations\com{in what hardware? using what software? Do we include source code? Otherwise it might not be relevant mentioning this}.
% This sounds inefficient but it is a trade off between time and memory.} \sxre{---Since just 2 sentences, and it's related to the last paragraph, I move it back to here. Or adding more and move to conclusion?---}\com{I'd say keep this as a comment in the conclusions if you'd like, but you'd need to be specific and I'd read it as implying that you're willing to share the code}

\section{Acknowledgement}
SXY thanks Ming-Chien Hsu and Shih-Min Yang for fruitful discussion.
SXY and PFR thank Mathias J\o{}rgensen for helpful comments and suggestions. The authors thank Dario Poletti and Guo Chu for insightful discussion.

\glsaddall
\printglossaries

\onecolumngrid
\newpage
\appendix
\glsresetall
\section{Randomized benchmarking for non-Markovian noise}
\subsection{The randomized benchmarking protocol}\label{Appendix: rb protocol}
A standard \gls{rb} experimental protocol proceeds as follows:
\begin{enumerate}
    \item Prepare an initial state $\rho_\syst$ on the system of interest $\syst$.
    \item Sample $m$ distinct elements, $\mc{G}_1,\mc{G}_2,\ldots,\mc{G}_m$, uniformly at random from a given gate set $\mbb{G}$. Let $\mc{G}_{m+1} := \Mcirc_{i=m}^1\mc{G}_i^\dg=\mc{G}_1^\dg\circ\cdots\circ\mc{G}_m^\dg$, where $\circ$ denotes composition of maps, and $\mc{G}^\dg(\cdot)=G^\dg(\cdot)G$ for any Kraus representation with unitaries $G$ of the map $\mc{G}$. We refer to $\mc{G}_{m+1}$ as an undo-gate.
    \item Apply the composition $\Mcirc_{i=1}^{m+1}\mc{G}_i$ on $\rho_\syst$. In practice, this amounts to applying a noisy sequence $\mc{S}_m:=\Mcirc_{i=1}^{m+1}\tilde{\mc{G}}_i$ of length $m$ on $\rho_\syst$, where $\tilde{\mc{G}}_i$ are the physical noisy gates associated to $\mc{G}$.
    \item Estimate the probability $f_m := \tr\left[M{\mc{S}}_m\left(\rho\right)\right]$ via a \gls{povm} element $M$.
    \item Repeat $n$ times the steps 1 to 4 for the same initial state $\rho_\syst$, same \gls{povm}~element $M$, and different sets of gates chosen uniformly at random $\{\mc{G}_i^{(1)}\}_{i=1}^m,\{\mc{G}_i^{(2)}\}_{i=1}^m,\ldots,\{\mc{G}_i^{(n)}\}_{i=1}^m$ from $\mbb{G}$ to obtain the probabilities $f_m^{(1)},f_m^{(2)},\ldots{f}_m^{(n)}$. Compute the \gls{asf}, $\mc{F}_m=1/n\sum_{i=1}^nf_m^{(i)}$.
    \item Examine the behavior of the \gls{asf}, $\mc{F}_m$, over different sequence lengths $m$.
\end{enumerate}

\subsection{The process tensor and non-Markovianity}\label{Appendix: pt and nm}

\section{The noise process tensor (model) and the control tensor (input)}\label{Appendix: tensors}
We want to compute the \gls{asf} for a sequence of $m+1$ noisy gates with an initial state $\rho_\syst$ and a \gls{povm} element ${M}$, written as a contraction of tensors
\begin{align}
    \mc{F}_m &= \mbb{E}\left[\Upsilon_{m+1} \ast \Theta_{m+1}(\rho_\syst,{M})\right] \\
    &= \mbb{E}\big\{\sum_{\vec{s}_{m+1}^{\,\,(\prime)},\,\vec{\zeta}_{m+1}^{\,\,(\prime)}}\Upsilon_{\vec{s}_{m+1}^{\,\,(\prime)}\vec{\zeta}_{m+1}^{\,\,(\prime)}}\, \Theta_{\vec{s}_{m+1}^{\,\,(\prime)}\vec{\zeta}_{m+1}^{\,\,(\prime)}}\big\}.
    \label{eq: asf tensor general}
\end{align}
where $\mbb{E}$ is average over gates, and $\ast$ here means an interior product
\begin{align}
    A\ast{B}=\tr[AB^\dg],
    \label{eq: def asterisk}
\end{align}
of the tensors $A$ and $B$, and $\vec{s}_{i}^{\,\,(\prime)}=(s_0,s_0^\prime,s_1,s_1^\prime,\ldots,s_{i},s_{i}^\prime)$, and similarly for $\vec{\zeta}^{\,\,(\prime)}_i$. Crucially, only the tensor $\Upsilon$ contains information about the noise and the environment $\env$.

We can identify the \gls{mpo} form of the tensors $\Upsilon$ and $\Theta$ by writing
\begin{equation}
    \mc{F}_m = \tr\{\mscr{Y}_{m+1}\,\mbb{E}(\mscr{G}_{m+1}^\mathrm{T})\},
\end{equation}
where $\mscr{Y}$ is the Choi state of the noise process and $\mscr{G}$ is the Choi state of the gate sequence, which is subsequently averaged ($\mbb{E}$) over gates; explicitly, these are defined by
\begin{equation}
    \mscr{Y}_{m+1} := \tr_\env\left[\Lambda_{m+1}\circ\mscr{S}_{m+1}\circ\cdots\circ\Lambda_1\circ\mscr{S}_1\circ\Lambda_0\left(\rho_\env\otimes\rho_\syst\otimes\psi^{\otimes{m+1}}\right)\right],
    \label{eq: appendix process tensor}
\end{equation}
where $\Lambda_i$ is the gate independent and \gls{cp} noise map at step $i$ acting on $\syst\env$, $\psi=\sum_{i,j=1}^{\dimS}|ii\rangle\!\langle{jj}|$ is an unnormalized maximally entangled state, with each copy defined on auxiliary spaces $\mathsf{A}_1\mathsf{B}_1,\ldots,\mathsf{A}_{m+1}\mathsf{B}_{m+1}$, $\mscr{S}_i$ is a swap operator between $\syst$ and half of the $i$\textsuperscript{th} auxiliary space, say $\mathsf{A}_i$, and $\rho_\env$ is a fiducial state of $\env$; on the other hand,
\begin{align}
    \mscr{G}_{m+1} &= {M}^\mathrm{T}\otimes\left(\mbb1_{\mathsf{A}_1}\otimes\mc{G}_1\otimes\mbb1_{\mathsf{A}_2}\otimes\mc{G}_2\otimes\cdots\otimes\mbb1_{\mathsf{A}_{m+1}}\otimes\mc{G}_{m+1}\right)\psi^{\otimes{m+1}},
\end{align}
with $\mc{G}_i$ the applied gate at step $i$. This definition is different from that in the main text in Eq.~\eqref{eq: gates choi state} simply in that here we incorporate directly the measurement $M$. Notice that the gates $\mc{G}_i$ have to be defined on the complementary auxiliary spaces, $\mathsf{B}_i$, because the swaps in Eq.~\eqref{eq: appendix process tensor} were defined with resp to $\mathsf{A}_i$. Finally, keep in mind that $\mc{G}_{m+1}$ is the undo (sequence of inverses) gate.

The only difference between $\mscr{Y}$ and $\Upsilon$, and between $\mbb{E}[\mscr{G}^\mathrm{T}]$ and $\Theta$ is the dependence of $\rho_\syst$: we want all the inputs in a \gls{mpo} separate from one containing solely the noise. For this reason, henceforth we refer to $\Upsilon$ as the noise tensor (or \gls{mpo}) and to $\Theta$ as the control tensor (or \gls{mpo}).

Consider $m=1$. Denote $\rho:=\Lambda_0(\rho_\env\otimes\rho_\syst)$ and let $\mathfrak{S}_{\alpha\beta}:=\mbb1_\env\otimes|\alpha\rangle\!\langle\beta|$, then
\begin{align}
    &\mscr{Y}_2 := \tr_\env[\Lambda_2\circ\mscr{S}_2\circ\Lambda_1\circ\mscr{S}_1\circ(\rho\otimes\psi^{\otimes2})] \nonumber\\
    &= \sum \tr_\env[\lambda_2^\mu\mathfrak{S}_{\alpha_2\beta_2}\lambda_1^\nu\mathfrak{S}_{\alpha_1\beta_1}\rho\mathfrak{S}_{\delta_1\gamma_1}\lambda_1^{\dg\,\nu}\mathfrak{S}_{\delta_2\gamma_2}\lambda_2^{\dg\,\mu}] \otimes |\beta_1\rangle\!\langle\alpha_1|i_1\rangle\!\langle{j_1}|\gamma_1\rangle\!\langle\delta_1|\otimes|i_1\rangle\!\langle{j}_1|\otimes|\beta_2\rangle\!\langle\alpha_2|i_2\rangle\!\langle{j_2}|\gamma_2\rangle\!\langle\delta_2|\otimes|i_2\rangle\!\langle{j}_2| \nonumber\\
    &= \sum \tr_\env[\lambda_2^\mu\mathfrak{S}_{\alpha_2\beta_2}\lambda_1^\nu\mathfrak{S}_{\alpha_1\beta_1}\rho\mathfrak{S}_{\delta_1\gamma_1}\lambda_1^{\dg\,\nu}\mathfrak{S}_{\delta_2\gamma_2}\lambda_2^{\dg\,\mu}]\otimes|\beta_1\alpha_1\beta_2\alpha_2\rangle\!\langle\delta_1\gamma_1\delta_2\gamma_2|, \nonumber\\
    &= \sum(\langle{e}|\otimes\mbb1_\syst)\lambda_2^\mu|e_2\alpha_2\rangle\!\langle{e}_2\beta_2|\lambda_1^\nu|e_1\alpha_1\rangle\!\langle{e}_1\beta_1|\rho|\epsilon_1\delta_1\rangle\!\langle\epsilon_1\gamma_1|\lambda_1^{\dg\,\nu}|\epsilon_2\delta_2\rangle\!\langle\epsilon_2\gamma_2|\lambda_2^{\dg\,\mu}(|e\rangle\otimes\mbb1_\syst) \otimes |\beta_1\alpha_1\beta_2\alpha_2\rangle\!\langle\delta_1\gamma_1\delta_2\gamma_2|
\end{align}
where in the last line we inserted complete $\env$ orthonormal bases, and $\lambda_i^\mu$ are Kraus operators of $\Lambda_i$. Onward, we will denote components $\syst\env$ of any $X$ matrix by
\begin{align}
    X_{ss^\prime}^{ee^\prime}:=\langle{e}s|X|e^\prime{s}^\prime\rangle.
\end{align}

Meanwhile,
\begin{align}
    \mscr{G}_2 &= {M}^\mathrm{T}\otimes\left(\mbb1_{\mathsf{A}_1}\otimes\mc{G}\otimes\mbb1_{\mathsf{A}_2}\otimes\mc{G}^\dg\right)\psi^{\otimes{2}} \nonumber\\
    &=M^\mathrm{T}\otimes\sum |i_1\rangle\!\langle{j}_1|\otimes\mc{G}(|i_1\rangle\!\langle{j}_1|)\otimes|i_2\rangle\!\langle{j}_2|\otimes\tilde{\mc{G}}(|i_2\rangle\!\langle{j}_2|),
\end{align}
where here $\tilde{\mc{G}}:= \mc{G}^{-1}$ is the undo-gate, and thus
\begin{align}
    &\tr[\mscr{Y}_2\mscr{G}_2^\transp] = \sum(\lambda_2^\mu)_{s\alpha_2}^{ee_2}(\lambda_1^\nu)_{\beta_2\alpha_1}^{e_2e_1}\rho_{\beta_1\delta_1}^{e_1\epsilon_1}(\lambda_1^{\dg\,\nu})^{\epsilon_1\epsilon_2}_{\gamma_1\delta_2}(\lambda_2^{\dg\,\mu})^{\epsilon_2e}_{\gamma_2s^\prime}\langle{s}^\prime|{M}|s\rangle \nonumber\\
    &\hspace{10em} \langle\delta_1\gamma_1\delta_2\gamma_2|\left\{|j_1\rangle\!\langle{i}_1|\otimes\mc{G}(|i_1\rangle\!\langle{j}_1|)^\transp\otimes|j_2\rangle\!\langle{i_2}|\otimes\tilde{\mc{G}}(|i_2\rangle\!\langle{j}_2|)^\transp\right\}|\beta_1\alpha_1\beta_2\alpha_2\rangle \nonumber\\
    &:=\sum (\lambda_2^\mu)_{s\alpha_2}^{ee_2}(\lambda_1^\nu)_{\beta_2\alpha_1}^{e_2e_1}\rho_{\beta_1\delta_1}^{e_1\epsilon_1}(\lambda_1^{\dg\,\nu})^{\epsilon_1\epsilon_2}_{\gamma_1\delta_2}(\lambda_2^{\dg\,\mu})^{\epsilon_2e}_{\gamma_2s^\prime} {M}_{s^\prime{s}}\,\langle\alpha_1|\mc{G}(|\beta_1\rangle\!\langle\delta_1|)|\gamma_1\rangle\!\langle\alpha_2|\tilde{\mc{G}}(|\beta_2\rangle\!\langle\delta_2|)|\gamma_2\rangle,
\end{align}

Let us now replace $\rho = \Lambda_0(\rho_\env\otimes\rho_\syst)$, and let $G$ be a unitary such that $\mc{G}(\cdot):=G(\cdot)G^\dg$ and similarly, in order to track the undo-gate, let $\tilde{G}$ such that $\tilde{\mc{G}}:=\tilde{G}(\cdot)\tilde{G}^\dg$; also let us relabel $\syst$ indices to a more consistent notation with labels $s$ and $\zeta$ (with all contractions remaining the same), so that
\begin{align}
    &\tr[\mscr{Y}_2\mscr{G}_2^\transp] = \sum (\lambda_2^\mu)_{s_2s_2^\prime}^{ee_2}(\lambda_1^\nu)_{s_1s_1^\prime}^{e_2e_1}\rho_{s_0\zeta_0^\prime}^{e_1\epsilon_1}(\lambda_1^{\dg\,\nu})_{\zeta_1 \zeta_1^\prime}^{\epsilon_1\epsilon_2}(\lambda_2^{\dg\,\mu})_{\zeta_2 \zeta_2^\prime}^{\epsilon_2e} {M}_{\zeta_2^\prime s_2} \,\langle{s}_1^\prime|\mc{G}(|s_0\rangle\!\langle\zeta_0^\prime|)|\zeta_1\rangle\!\langle{s_2^\prime}|\tilde{\mc{G}}(|s_1\rangle\!\langle\zeta_1^\prime|)|\zeta_2\rangle \nonumber\\
    &= \sum (\lambda_2^\mu)_{s_2s_2^\prime}^{ee_2}(\lambda_1^\nu)_{s_1s_1^\prime}^{e_2e_1}(\lambda_0^\sigma)_{s_0s_0^\prime}^{e_1e_0}(\rho_\env)^{e_0\epsilon_0}(\rho_\syst)_{s_0^\prime \zeta_0}(\lambda_0^{\dg\,\sigma})_{\zeta_0\zeta_0^\prime}^{\epsilon_0\epsilon_1}(\lambda_1^{\dg\,\nu})_{\zeta_1 \zeta_1^\prime}^{\epsilon_1\epsilon_2}(\lambda_2^{\dg\,\mu})_{\zeta_2 \zeta_2^\prime}^{\epsilon_2e} {M}_{\zeta_2^\prime s_2} G_{{s}_1^\prime s_0}G^\dg_{\zeta_0^\prime\zeta_1}\tilde{G}_{s_2^\prime s_1}\tilde{G}^\dg_{\zeta_1^\prime\zeta_2},
\end{align}
so we may now identify
\begin{align}
    \Upsilon_{\vec{s}_2^{\,\,(\prime)}\vec{\zeta}_2^{\,\,(\prime)}} &= \sum_{\mu,\nu,\sigma}\sum_{e_i,\epsilon_i} (\lambda_2^\mu)_{s_2s_2^\prime}^{ee_2}(\lambda_1^\nu)_{s_1s_1^\prime}^{e_2e_1}(\lambda_0^\sigma)_{s_0s_0^\prime}^{e_1e_0}(\rho_\env)^{e_0\epsilon_0}(\lambda_0^{\dg\,\sigma})_{\zeta_0\zeta_0^\prime}^{\epsilon_0\epsilon_1}(\lambda_1^{\dg\,\nu})_{\zeta_1 \zeta_1^\prime}^{\epsilon_1\epsilon_2}(\lambda_2^{\dg\,\mu})_{\zeta_2 \zeta_2^\prime}^{\epsilon_2e},\\
    \Theta_{\vec{s}_2^{\,\,(\prime)}\vec{\zeta}_2^{\,\,(\prime)}}
    &= \tilde{G}_{s_2^\prime{s}_1}\,G_{s_1^\prime{s}_0} \, (\rho_\syst)_{s_0^\prime \zeta_0} \,G^\dg_{\zeta_0^\prime\zeta_1}\, \tilde{G}^\dg_{\zeta_1^\prime\zeta_2}\,{M}_{\zeta_2^\prime s_2} ,
\end{align}
so generally for any $m\geq1$,
\begin{align}
    \Upsilon_{\vec{s}_{m+1}^{\,\,(\prime)}\vec{\zeta}_{m+1}^{\,\,(\prime)}} &= \sum_{\mu_i}\sum_{\{e,\epsilon\}} (\lambda_{m+1}^{\mu_{m+1}})_{s_{m+1}s_{m+1}^\prime}^{ee_{m+1}}\left[\prod_{i=0}^m(\lambda_i^{\mu_i})_{s_is_i^\prime}^{e_{i+1}e_i}\right](\rho_\env)^{e_0\epsilon_0}\left[\prod_{j=0}^m(\lambda_j^{\dg\,\mu_j})_{\zeta_j\zeta_j^\prime}^{\epsilon_j\epsilon_{j+1}}\right](\lambda_{m+1}^{\dg\,\mu_{m+1}})_{\zeta_{m+1} \zeta_{m+1}^\prime}^{\epsilon_{m+1}e},
    \label{eq: noise tensor}\\
    \Theta_{\vec{s}_{m+1}^{\,\,(\prime)}\vec{\zeta}_{m+1}^{\,\,(\prime)}} &= \tilde{G}_{s_{m+1}^\prime s_m}\left[\prod_{i=1}^m (G_i)_{s_i^\prime s_{i-1}}\right](\rho_\syst)_{s_0^\prime \zeta_0}\left[\prod_{j=1}^m (G^\dg_j)_{\zeta_{j-1}^\prime\zeta_j}\right]\tilde{G}^\dg_{\zeta_m^\prime\zeta_{m+1}}\,{M}_{\zeta_{m+1}^\prime s_{m+1}},
    \label{eq: control tensor}
\end{align}
where here $\mc{G}_i:=G_i(\cdot)G_i^\dg$ is the gate at the $i$\textsuperscript{th} step for corresponding unitaries $G_i$, and $\tilde{G}:=\left(G_{m}G_{m-1}\cdots{G}_1\right)^\dg$ gives the undo-gate $\tilde{\mc{G}}=\tilde{G}(\cdot)\tilde{G}^\dg$. These tensors are depicted in Fig.~\ref{fig: upsilon omega} in the main text. Both tensors are of rank $4(m+2)$, i.e., each has $4(m+2)$ legs of dimension $\dimS$, and so there is a total of $\dimS^{\,4(m+2)}$ components to be specified on each.

Notice that if the $\Lambda_i$ matrices are unitary (so that all $\lambda_i^{\mu_i}$ have only a single $\mu_i$), then for $\Upsilon$ we require a memory scaling as $\dimE^2\left[1+(m+2)\dimS^2\right]$. Similarly, for $\Theta$ we require $\left(m+2\right)\dimS^2$ parameters. That is, we specify the nodes $\Lambda_i$, $\rho_\env$, $\mc{G}_i$, $\rho_\syst$ and $M$ in each tensor to obtain either $\Upsilon$ or $\Theta$. Furthermore, assuming the size of the environment is sufficiently large, the effective bond dimension in $\Upsilon$ between timesteps need not necessarily be $\dimE$ but often will be smaller, giving rise to the effective noise dynamics specified by corresponding effective \gls{cp} maps. Finally, notice that we may get rid of the complex conjugate in the final expressions and simply compute conjugates of entries. We thus would have, for unitary noise,
\begin{align}
    \Upsilon_{\vec{s}_{m+1}^{\,\,(\prime)}\vec{\zeta}_{m+1}^{\,\,(\prime)}} &= \sum_{\{e,\epsilon\}} (\lambda_{m+1})_{s_{m+1}s_{m+1}^\prime}^{ee_{m+1}}\left[\prod_{i=0}^m(\lambda_i)_{s_is_i^\prime}^{e_{i+1}e_i}\right](\rho_\env)^{e_0\epsilon_0}\left[\prod_{j=0}^m(\lambda_j)_{\zeta_j^\prime\zeta_j}^{*\,\epsilon_{j+1}\epsilon_j}\right](\lambda_{m+1})_{\zeta_{m+1}^\prime\zeta_{m+1}}^{*e\epsilon_{m+1}},
    \label{eq: noise tensor unitary}\\
    \Theta_{\vec{s}_{m+1}^{\,\,(\prime)}\vec{\zeta}_{m+1}^{\,\,(\prime)}} &= \mbb{E}\left\{\hat{G}^*_{s_m s_{m+1}^\prime}\left[\prod_{i=1}^m (G_i)_{s_i^\prime s_{i-1}}\right](\rho_\syst)_{s_0^\prime \zeta_0}\left[\prod_{j=1}^m (G_j)^*_{\zeta_j\zeta_{j-1}^\prime}\right]\hat{G}_{\zeta_m^\prime\zeta_{m+1}}\,{M}_{\zeta_{m+1}^\prime s_{m+1}}\right\},
    \label{eq: control tensor unitary}
\end{align}
where here $*$ denotes complex conjugate, and $\hat{G}:=G_{m}G_{m-1}\cdots{G}_1$.

A particular tensor that we will care about, is that obtained by a partial contraction of $\Upsilon_m$ and $\Theta_m$ whereby given nodes are not contracted. That is, denoting by $\Upsilon_{m}^{\{i,j,\ldots,k\}}$ a noise tensor without the $\lambda_i^{\mu_i}$, $\lambda_j^{\mu_j}$, \ldots, $\lambda_k^{\mu_k}$ nodes (i.e., with such $\lambda$ terms not appearing in Eq.~\eqref{eq: noise tensor}), we define
\begin{align}
    \Tilde{\Theta}_{m+1}^{\{i,j,\ldots,k\}} := \sum\Upsilon^{\{i,j,\ldots,k\}}_{\vec{s}_{m+1}^{\,(\prime)}\vec{\zeta}_{m+1}^{\,(\prime)}}\Theta_{\vec{s}_{m+1}^{\,(\prime)}\vec{\zeta}_{m+1}^{\,(\prime)}},
    \label{eq: tilde theta def}
\end{align} as the contraction of all the tensors in the \gls{asf} of sequence length $m$, except for $\lambda_i$, $\lambda_j$, \ldots, $\lambda_k$.
%--------------------------------------------------
% , \add{as illustrated in Fig.~\com{XXXX}}.
%--------------------------------------------------

\subsection{The (multi-qubit) Clifford group case}\label{Appendix: rb clifford nM}
In~\cite{figueroaromero2021randomised}, the average $\mbb{E}[\mscr{G}^\dg] = \mbb{E}[\mscr{G}^\mathrm{T}]$ was analytically evaluated when the gates belong to the (multi-qubit) Clifford group, or more generally to any unitary 2-design, and the contraction with $\mscr{Y}$ was done explicitly to obtain the \gls{asf} in Eq.~\eqref{eq: asf tensor general}, which can be written as
\begin{align}
    \mc{F}_m &:=  \tr\left[M\,\tr_\env\circ\Lambda_{m+1}\circ\left(\mscr{A}_m+\mscr{B}_m\right)\rho\right],
    \label{eq: average fidelity nonMarkovian}
\end{align}
where
\begin{align}
    \mscr{A}_m(\rho) &:= \f{\displaystyle{\Mcirc_{n=1}^m}\left(\$_{\Lambda_n}-\pounds_{\Lambda_n}\right)\otimes\mc{I}_\syst}{\left(\dimS^2-1\right)^m}\left(\rho-\tilde{\rho}_\env\otimes\f{\mbb1}{\dimS}\right), \label{eq: main curlyA}\\
    \mscr{B}_m(\rho) &:= \left(\Mcirc_{n=1}^m\pounds_{\Lambda_n}\right)\tilde{\rho}_\env\otimes\f{\mbb1}{\dimS}, \label{eq: main curlyB}
\end{align}
with $\rho=\Lambda_0(\rho_\env\otimes\rho_\syst)$ and with $\tilde{\rho}_\env:=\tr_\syst[\Lambda_0(\rho_\env\otimes\rho_\syst)]$; here $\$_{\Lambda_n},\pounds_{\Lambda_n}$ are maps acting solely on $\env$ as defined by
\begin{align}
    \$_{\Lambda_n}(\varepsilon) &:=\sum_{s,s^\prime=1}^{\dimS} \langle{s}|\Lambda_n(\varepsilon\otimes|s\rangle\!\langle{s^\prime}|)|s^\prime\rangle \label{eq: dollar main} \\
    \pounds_{\Lambda_n}(\varepsilon) &:= \tr_\syst\left[\Lambda_n\left(\varepsilon\otimes\f{\mbb1}{\dimS}\right)\right] \label{eq: pounds main},
 \end{align}
for any given operator $\varepsilon$ acting on $\env$. This is seen to imply that the tensor $\Theta_{m+1}$ is similar to the elements of a depolarizing tensor on $\syst$, with components
\begin{align}
    &\Theta_{\vec{s}_{m+1}^{\,(\prime)}\vec{\zeta}_{m+1}^{\,(\prime)}} = \left[\alpha_{\vec{s}_{m+1}^{\,(\prime)}\vec{\zeta}_{m+1}^{\,(\prime)}} + \beta_{\vec{s}_{m+1}^{\,(\prime)}\vec{\zeta}_{m+1}^{\,(\prime)}}  \right](\rho_\syst)_{s_0^\prime\zeta_0}M_{\zeta^\prime_{m+1}s_{m+1}},
    \label{eq: theta 2-designs}
\end{align}
where here
\begin{align}
    \alpha_{\vec{s}_{m+1}^{\,(\prime)}\vec{\zeta}_{m+1}^{\,(\prime)}} &:= \f{\displaystyle{\prod_{n=1}^m}\left(\dimS\delta_{s_ns^\prime_n}\delta_{\zeta^\prime_n\zeta_n}-\delta_{s_n\zeta^\prime_n}\delta_{s^\prime_n\zeta_n}\right)}{\dimS^m(\dimS^2-1)^m}\left(\delta_{s_0s_{m+1}^\prime}\delta_{\zeta_0^\prime\zeta_{m+1}} - \f{\delta_{s_0\zeta_0^\prime}\delta_{s^\prime_{m+1}\zeta_{m+1}}}{\dimS}\right) \\
    \beta_{\vec{s}_{m+1}^{\,(\prime)}\vec{\zeta}_{m+1}^{\,(\prime)}} &:= \f{\displaystyle{\prod_{n=1}^m}\delta_{s_n\zeta^\prime_n}\delta_{s_n^\prime\zeta_n}}{\dimS^m}\left(\f{\delta_{s_0\zeta_0^\prime}\delta_{s^\prime_{m+1}\zeta_{m+1}}}{\dimS}\right)
\end{align}
are the terms analogous to the $p$ and $1-p$ terms in the depolarizing channel. Notice that these definitions are slightly different to those made in~\cite{figueroaromero2021randomised}; this is simply because of the definition $\rho = \Lambda_0(\rho_\env\otimes\rho_\syst)$ and because $\Theta_m$ takes $\rho_\syst$ while $\Upsilon_m$ takes $\lambda_0^{(\dg)}$ and $\rho_\env$.

\section{Sweeping algorithm for Randomized Benchmarking}\label{Appendix: RB sweeping algorithm}
The so-called sweeping algorithm from~\cite{stoudenmire2016} combines the idea of the kernel trick in~\cite{kernel_trick} and the density matrix renormalization group (DMRG) algorithm in~\cite{DMRG}: it is a tensor network supervised learning model which predicts outputs via the contraction of parameter tensor and input tensor.
In a \gls{rb} experiment, we obtain an experimental \gls{asf} and we choose the control tensor, but the underlying noise tensor generally remains unknown. Here we adapt the sweeping algorithm to obtain information about the average noise involved within a \gls{rb} experiment.
Taking the control tensor as an input tensor, the noise tensor as parameter tensor, and the experimental \gls{asf} as the label, we  optimize the model until the predicted \gls{asf} is close enough (in a sense to be made precise below) to the experimental \gls{asf}.

More specifically, let us consider the noise tensor $\Upsilon_m$ written as the \gls{mpo} in Eq.~\eqref{eq: noise tensor}, for any fixed $m\geq1$. Our goal is now to implement an extension of the sweeping algorithm above to learn features about $\Upsilon_m$, namely some model of the $\Lambda$ noise maps.
% and ultimately the required bond-dimension of the noise tensor that approximately reproduces the $m$ points of some given experimental set of \gls{asf}s, $\{\asfexp_n\}_{n=1}^m$. 
While the corresponding \gls{rb} experiment may be carried out with an arbitrary gate set, the closed expression in~\cite{figueroaromero2021randomised} for the Clifford group, can make the evaluation of the corresponding learned {\gls{asf}}, $\mc{F}_m$, in exact form readily available.

As opposed to the general sweeping algorithm, the noise tensor $\Upsilon_m$ has physical meaning, and thus it has certain constraints to follow.
In particular, we fix all $\Lambda$ maps to be unitary.
% as the algorithm will be able to tell us whether some environmental dimensions are redundant, thus reducing such maps to \gls{cp} maps on a smaller effective environment; the tradeoff is that we have to choose a \emph{sufficiently large} environment, which is a-priori unknown. Nevertheless, 
Imposing this unitarity constraint will allow us to sweep over only half the nodes of the noise tensor, thereby obtaining the rest by taking the adjoint of the updated nodes.

Let us then begin by defining a cost function
\begin{align}
    C &:= \frac{1}{2}\sum_{n=1}^m\left(\mc{F}_n - \asfexp_n\right)^2,
\end{align}
where as mentioned above, $\asfexp_1$, $\asfexp_2$, \ldots, $\asfexp_m$, denote experimental \gls{asf} points for a \gls{rb} sequence with up to length $m$, and $\mc{F}_1$, $\mc{F}_2$, \ldots, $\mc{F}_m$ denote the learned points by the algorithm. The goal is then for the algorithm to determine noise maps $\Lambda$ that minimize $C$.

We now denote by $N_\env$ and $N_\syst$ the number of qubits on the environment and system, respectively, so that $\dimE:=2^{N_\env}$ and $\dimS:=2^{N_\syst}$. The algorithm will now proceed as described below.
For fear that the fluency of the description will be break by the detail of constrained optimization, regarding Step~\ref{alg:contr_opt}, we will discuss it in Appendix~\ref{Appendix: u_constraint}.
\begin{enumerate}
    \item\label{alg:initialize} Set an appropriate $N_\env$, as discussed in the main text, and require all $\Lambda$ maps to be unitary, i.e., $\Lambda_i(\cdot):=\lambda_i(\cdot)\lambda_i^\dg$ such that $\lambda_i\lambda_i^\dg=\lambda_i^\dg\lambda_i=\mbb1_{\env\syst}$.  Initialize each node $\lambda_i$ in the noise \gls{mpo} $\Upsilon_m$; this choice is arbitrary, but here we employ identity matrices (i.e., we start assuming an absence of noise). Finally, set the initial state of the environment $\rho_\env=|0\rangle\!\langle0|$; this choice is also arbitrary and fixed throughout the algorithm: any information about how this fiducial state changes is to be contained in the state preparation noise $\Lambda_0$.
    
    \item\label{algo:bonding} Take any two neighboring nodes 
    of the noise \gls{mpo}, $\lambda_i$ and $\lambda_{i-1}$ for any $1\leq{i}\leq{m+1}$, and contract them on $\env$ into a joint node $L_{i,i-1}$, i.e., \begin{equation}
        (L_{i,i-1})^{e_{i+1}e_{i-1}}_{s_{i}s_{i}^{\prime}s_{i-1}s_{i-1}^\prime} := \sum_{e_{i}} (\lambda_{i})_{s_{i}s_{i}^\prime}^{e_{i+1}e_{i}} (\lambda_{i-1})_{s_{i-1}s_{i-1}^\prime}^{e_{i}e_{i-1}},
    \end{equation}
    such that
    \begin{align}     \Upsilon_{\vec{s}_{m+1}^{\,\,(\prime)}\vec{\zeta}_{m+1}^{\,\,(\prime)}} 
    &= \sum_{\{e,\epsilon\}/e_{i}}\cdots
    \left[ \sum_{e_{i}} (\lambda_{i})_{s_{i}s_{i}^\prime}^{e_{i+1}e_{i}} (\lambda_{i-1})_{s_{i-1}s_{i-1}^\prime}^{e_{i}e_{i-1}}\right]\cdots
   (\lambda_0)_{s_0s_0^\prime}^{e_1e_0}
    (\rho_\env)^{e_0\epsilon_0}\cdots\nonumber\\
    &= \sum_{\{e,\epsilon\}/e_{i}}\cdots \left[(L_{i,i-1})^{e_{i+1}e_{i-1}}_{s_{i}s_{i}^{\prime}s_{i-1}s_{i-1}^\prime}\right] \cdots(\lambda_0)_{s_0s_0^\prime}^{e_1e_0} (\rho_\env)^{e_0\epsilon_0}\cdots.
    \end{align}
%--------------------------------------------------
    % \add{as illustrated in Fig.~\com{XXXX}.}
%--------------------------------------------------

    The joint node $L_{i,i-1}$ contains the parameters to be updated, and the algorithm will then similarly \emph{sweep} through all the remaining bonds. While the sweeping order is in principle arbitrary, it can preferably be made in an orderly fashion, either ascending or descending.
        
    \item\label{alg:grad} Take the gradient of the cost function with respect to the joint node:
    \begin{align}
    \Delta L_{i,i-1} &= -\frac{\partial C}{\partial L_{i,i-1}} \nonumber\\
    &= \sum_{n=1}^m \left( \asfexp_n - \mc{F}_n\right)\frac{\partial \mc{F}_n}{\partial L_{i,i-1}}\nonumber\\
    &= \sum_{n=i-1\,;\,n\geq1}^m \left(\asfexp_n - \mc{F}_n\right) \frac{\partial \mc{F}_n}{\partial L_{i,i-1}},
    \end{align}
    where for $n \leq i-2$, $\partial \mc{F}_{n}/\partial L_{i:i-1}=0$, since these $\mc{F}_n$ terms do not explicitly depend on the joint node, $L_{i,i-1}$.
    The corresponding derivative is given by
    \begin{align}
    \frac{\partial \mc{F}_n}{\partial \left(L_{i,i-1}\right)_{s_i s_i^\prime s_{i-1}s_{i-1}^\prime}^{e_{i+1}e_{i-1}}} &= \frac{\partial\left(L_{i,i-1} \ast \Tilde{\Theta}_{n+1}^{\{i,i-1\}}\right)}{\partial \left(L_{i,i-1}\right)_{s_i s_i^\prime s_{i-1}s_{i-1}^\prime}^{e_{i+1}e_{i-1}}} = \left(\Tilde{\Theta}_{n+1}^{\{i,i-1\}}\right)_{s_i s_i^\prime s_{i-1}s_{i-1}^\prime}^{e_{i+1}e_{i-1}},
        \end{align}
        where the tensor $\tilde\Theta_{n}^{\{i,j,\ldots,k\}}$ is defined in Eq.~\eqref{eq: tilde theta def} as the contraction of all tensors in the \gls{asf}, except $\lambda_i$, $\lambda_j$, \ldots, $\lambda_k$. Now, the components of the gradient become
        \begin{align}
            (\Delta L_{i,i-1} &)^{e_{i+1}e_{i-1}}_{s_is_i^\prime s_{i-1} s_{i-1}^\prime} =
            \sum_{n=i-1\,:\,n\geq1}^m \left(\asfexp_n - \mc{F}_n\right) \left(\Tilde{\Theta}_{n+1}^{\{i,i-1\}}\right)_{s_i s_i^\prime s_{i-1}s_{i-1}^\prime}^{e_{i+1}e_{i-1}}.
            \label{eq: delta L}
        \end{align}
        
        Each $\mc{F}_n$ here corresponds to the \gls{asf} with noise $\Lambda_1,\ldots,\Lambda_n$ at the current step, and it can be computed generally through Eq.~\eqref{eq: asf tensor general}. If the gates in question belong to a unitary 2-design (e.g. multi-qubit Clifford gates), the closed expression in Eq.~\eqref{eq: average fidelity nonMarkovian} can be used instead.
    \item\label{alg:updating} Define an updated joint node by defining,
        \begin{equation}
        L_{i,i-1}^\prime := L_{i,i-1} + \alpha\,\Delta L_{i,i-1},
        \end{equation}
        where $\alpha$ is a positive parameter controlling the scale of the update.
    \item\label{alg:svd} Split the joint node by a higher dimensional version of the \gls{svd}~\cite{Bridgeman_2017}. Grouping the indices as $A = \dimS^2 e_{i+1} + \dimS s_i + s_i^\prime$ and $B = \dimS^2 e_{i-1} + \dimS s_{i-1} + s_{i-1}^\prime$, and followed by the \gls{svd},
        \begin{align}
        (L_{i,i-1}^\prime)_{AB}
        &= \sum_{\epsilon=1}^{\dimE\dimS^2} U_{A\epsilon} S_{\epsilon\epsilon} (V^\dg)_{\epsilon B}\\
        &= \sum_{\epsilon=1}^{\dimE\dimS^2} (U\sqrt{S})_{A \epsilon} (\sqrt{S}V^\dg)_{\epsilon B}.
        \end{align}
    In order to maintain the original dimension of the nodes, truncate by inserting 
    $R = 
    \begin{bmatrix}
    \mbb1_{\dimE}\\
    0_{\dimE\dimS^2-\dimE, \dimE}
    \end{bmatrix}$ and 
    $R^T = 
    \begin{bmatrix}
    \mbb1_{\dimE} & 0_{\dimE, \dimE\dimS^2-\dimE}
    \end{bmatrix}$, where $0_{a,b}$ is a $a \times b$ null matrix.
        \begin{align}
        (L_{i,i-1}^\prime)_{AB}
        & \approx \sum_{\epsilon =1}^{\dimE\dimS^2} \sum_{\epsilon^\prime =1}^{\dimE} (U\sqrt{S})_{A\epsilon} R_{\epsilon \epsilon^\prime} (R^T)_{\epsilon^\prime \epsilon} (\sqrt{S}V^\dg)_{\epsilon B}\\
        &= \sum_{\epsilon^\prime =1}^{\dimE}
        \tilde{U}_{A \epsilon^\prime} (\tilde{V}^\dg)_{\epsilon^\prime B}.
        \end{align}
        Then split the indices back out,
        \begin{align}
        (L_{i,i-1}^\prime)^{e_{i+1}e_{i-1}}_{s_{i}s_{i}^{\prime}s_{i-1}s_{i-1}^\prime}
        &\approx \sum_{\epsilon^\prime =1}^{\dimE} \tilde{U}^{e_{i+1} \epsilon^\prime}_{s_is_i^\prime} (\tilde{V}^\dg)^{\epsilon^\prime e_{i-1}}_{s_{i-1}s_{i-1}^\prime},
        \end{align}
        where $\tilde{U}$ and $\tilde{V}^\dg$ are not necessarily unitary anymore.
        
    \item\label{alg:contr_opt}
        To project $\tilde{U}$ and $\tilde{V}$ into unitaries, first group $A_1= \dimS e_{i+1} + s_i$, $B_1=  \dimS \epsilon^\prime + s_i^\prime$ and $A_2= \dimS \epsilon^\prime + s_{i-1}$, $B_2= \dimS e_{i-1} + s_{i-1}^\prime$, and followed by the \gls{svd},
        \begin{align}
        \label{eq: truncated_u}
            \tilde{U}_{A_1B_1} 
            &= \sum_{\beta_1=1}^{\dimE\dimS} (U_1)_{A_1 \beta_1} (\Sigma_1)_{\beta_1 \beta_1} (V_1^\dg)_{\beta_1 B_1},\\
            (\tilde{V}^\dg)_{A_2B_2} 
            &= \sum_{\beta_2=1}^{\dimE\dimS} (U_2)_{A_2 \beta_2} (\Sigma_2)_{\beta_2 \beta_2} (V_2^\dg)_{\beta_2 B_2},
        \end{align}
        where $\Sigma_i$ is the diagonal matrix and $U_i$, $V_i$ are the unitary matrices with $i=1$ corresponding to $\tilde{U}$ and $i=2$ corresponding to $\tilde{V^\dg}$. Define the projections
        \begin{align}
        \label{eq: proj_u}
        \pi(\tilde{U}) &= U_1 V_1^\dg,\\
        \label{eq: proj_v}
        \pi(\tilde{V^\dg}) &= U_2 V_2^\dg.
        \end{align}
        
    \item\label{alg:replace} Replace all nodes
        \begin{equation}
            \lambda_n\to  \pi(\tilde{U})\pi(\tilde{V^\dg}),
        \end{equation}
        and the respective conjugates. This ensures unitarity and time-independence
    %--------------------------------------------------
        % ~\com{we might need to relax this and replace only each pair of nodes at a time; also, at least with how the decomposition is done, all bond dimensions would be the same if the model is time-independent, which is no good (in general)} 
    %--------------------------------------------------
        in the model.
        When this step is finished, one iteration is completed.
        
        \item Go back to step \ref{algo:bonding} taking a different arbitrary pair of neighboring updated nodes. Repeat until the predicted \gls{asf}, $\mc{F}_m$, is close enough to the experimental \gls{asf}, $\asfexp_m$, according to the $\ell_1$-norm,
        \begin{equation}
        \sum_{n=1}^m|\mc{F}_n - \asfexp_n| \leq \sigma_T/\delta,
        \end{equation}
        where $\delta\geq1$ and $\sigma_T$ is the sum of standard errors of the mean associated to each difference of \gls{asf} points, i.e. $\sigma_T=\sum_n\left|\sigma^{\mathrm{(exp)}}_n-\sigma_n\right|$, with individual $\sigma^{\mathrm{(exp)}}_n$ as given by the experiment, and $\sigma_n$ those corresponding to the sampled gates used for the learned points; for the Clifford case, if the exact average in Eq.~\eqref{eq: theta 2-designs} is employed, then all $\sigma_n=0$.
\end{enumerate}

\section{Unitary constraint optimization}
\label{Appendix: u_constraint}

For self-containment and reader's convenience, we directly quote here the context given in~\cite{984753}.

\textit{Definition 4 in~\cite{984753} (Stiefel Manifold):} The complex Stiefel manifold $S t(n, p)$ is the set
$$
S t(n, p)=\left\{X \in \mathbb{C}^{n \times p}: X^{H} X=\mbb1\right\},
$$
for $n\geq p$ and $X^H$ is the Hermitian conjugate of $X$. This is our unitary constraint set when $n=p$.

\textit{Definition 5 in~\cite{984753} (Projection):} Let $X \in \mathbb{C}^{n \times p}$ be a rank $p$ matrix. The projection operator $\pi: \mathbb{C}^{n \times p} \rightarrow S t(n, p)$ onto the Stiefel manifold $S t(n, p)$ is defined to be
$$
\pi(X)=\arg \min _{Q \in S t(n, p)}\|X-Q\|^{2} .
$$
The following useful lemma follows immediately from the fact that $\|U X V\|=\|X\|$ if $U$ and $V$ are unitary.

\textit{Lemma 6 in~\cite{984753}:} If $U \in \mathbb{C}^{n \times n}$ and $V \in \mathbb{C}^{p \times p}$ are unitary matrices, then $\pi(U X V)=U \pi(X) V$.

\textit{Proposition 7 in~\cite{984753}:} Let $X \in \mathbb{C}^{n \times p}$ be a rank $p$ matrix. Then, $\pi(X)$ is well defined. Moreover, if the SVD of $X$ is $X=$ $U \Sigma V^{H}$, then $\pi(X)=U \mbb1_{n, p} V^{H}$.

Unitary matrix has full rank, truncated would still have full truncated row/column rank. Thus, $\tilde{U}$ and $\tilde{V^\dg}$ in Eq.~\ref{eq: truncated_u} are full rank matrices.
From \textit{Lemma 6} and \textit{Proposition 7}, we claim Eq.~\ref{eq: proj_u} and Eq.~\ref{eq: proj_v} are the projection of $\tilde{U}$ and $\tilde{V^\dg}$.

% \section{Interpretation of outputs and scaling of the algorithm}
\section{Interpretation of outputs}
\label{Appendix: interpretation}

\subsection{Markovian noise model}
\label{Appendix: interpret_M}

The learned noise is an artificial matrix which reproduce a similar statistic of \gls{asf} data.
But it has an interesting feature when learned from a Markovian noise model.
The following is the learned Markovian noise rounding to $10^{-7}$ for the phase flip noise,
{\fontsize{10}{10} \selectfont $$
\begin{bmatrix}
0.977-0.002i& -0.2  -0.075i&  0   & 0\\
0.2  -0.075i&  0.977   &  0   &  0\\
0   & 0   &  1  &  0\\
0   & 0   & 0   &  1
\end{bmatrix}.
$$}
It corresponds to a direct sum of a unitary matrix and an identity. The upper left square is a unitary when checking the raw matrix,
{\fontsize{10}{10} \selectfont $$
\begin{bmatrix}
9.76909396e-01-1.88659124e-03i& -2.00029693e-01-7.50506084e-02i\\ 
1.99891456e-01-7.54180194e-02i&  9.76911214e-01+9.15991865e-05i
\end{bmatrix}.
$$}
To verify this is a common feature when learning from a Markovian noise model, we also check a Markovian amplitude damping noise model in a \gls{rb} protocol with sequence length $40$.
The noise from our method rounding to $10^{-5}$ is
{\fontsize{10}{10} \selectfont $$
\begin{bmatrix}
-0.8406 +0.52257i&  0.13549+0.04416i& 0     & 0\\
-0.02043+0.14103i& -0.84087+0.52214i&  0    & 0\\
0     &  0     &  1    & 0\\
0     & 0     & 0     & 1
\end{bmatrix}.
$$}
Again, from the raw matrix it is indeed an unitary,
{\fontsize{10}{10} \selectfont $$
\begin{bmatrix}
-8.40603627e-01+5.22568321e-01i&  1.35492168e-01+4.41561374e-02i\\
-2.04332866e-02+1.41033233e-01i& -8.40870789e-01+5.22138320e-01i
\end{bmatrix}.
$$}
Finally, we verified this with a depolarising noise model, which consistently displays the same feature.

\subsection{Non-Markovian noise model}
\label{Appendix: interpret_nM}

The noise unitary from the non-Markovian model in Section~\ref{subsec: learned nM} is
{\fontsize{10}{10} \selectfont $$
\begin{bmatrix}
 0.991 & 0.054-0.058i& 0.054-0.058i& -0. -0.066i\\
-0.061-0.058i& 0.991 & -0.007-0.06i & 0.054-0.058i\\
-0.061-0.058i& -0.007-0.06i & 0.991 & 0.054-0.058i\\
 0. -0.053i& -0.061-0.058i& -0.061-0.058i& 0.991
\end{bmatrix}.
$$}
The corresponding learned non-Markovian noise rounding to $10^{-3}$ is 
{\fontsize{10}{10} \selectfont $$
\begin{bmatrix}
0.997   & -0.078-0.021i& -0.016   &  0.002-0.001i\\
0.079-0.021i&  0.995   &  0.046   &  0.018-0.004i\\
0.012+0.001i& -0.048-0.001i&  0.999+0.002i&  0.009-0.001i\\
-0.004-0.001i& -0.018-0.003i& -0.009-0.001i&  1.   -0.002i
\end{bmatrix},
$$}
They can both be seen as an identity plus a small noise matrix.
The more precise matrix is provided here. It has too many digits to be displayed, but it is enough to check the unitarity up to $10^{-9}$ precision.
{\fontsize{10}{10} \selectfont $$
\begin{bmatrix}
 0.996582  -0.00013661i& -0.07812799-0.02128808i& -0.01616884-0.00028138i& 0.00205671-0.00117742i\\
0.07874099-0.02131937i& 0.99540601-0.00016128i& 0.04647442-0.00035591i& 0.01839634-0.00371055i\\
0.01242521+0.00069801i& -0.04775881-0.00077703i& 0.99874221+0.0019062i & 0.00856528-0.00078986i\\ -0.00368633-0.00108177i& -0.01777916-0.00347567i& -0.00938026-0.00093297i&  0.99978209-0.00199489i
\end{bmatrix}.
$$}

\subsection{The sudden jump in the cost function}
\label{Appendix: interpret_cost}
\begin{figure}[h!]
\begin{minipage}{.47\textwidth}
    \includegraphics[width=\linewidth]{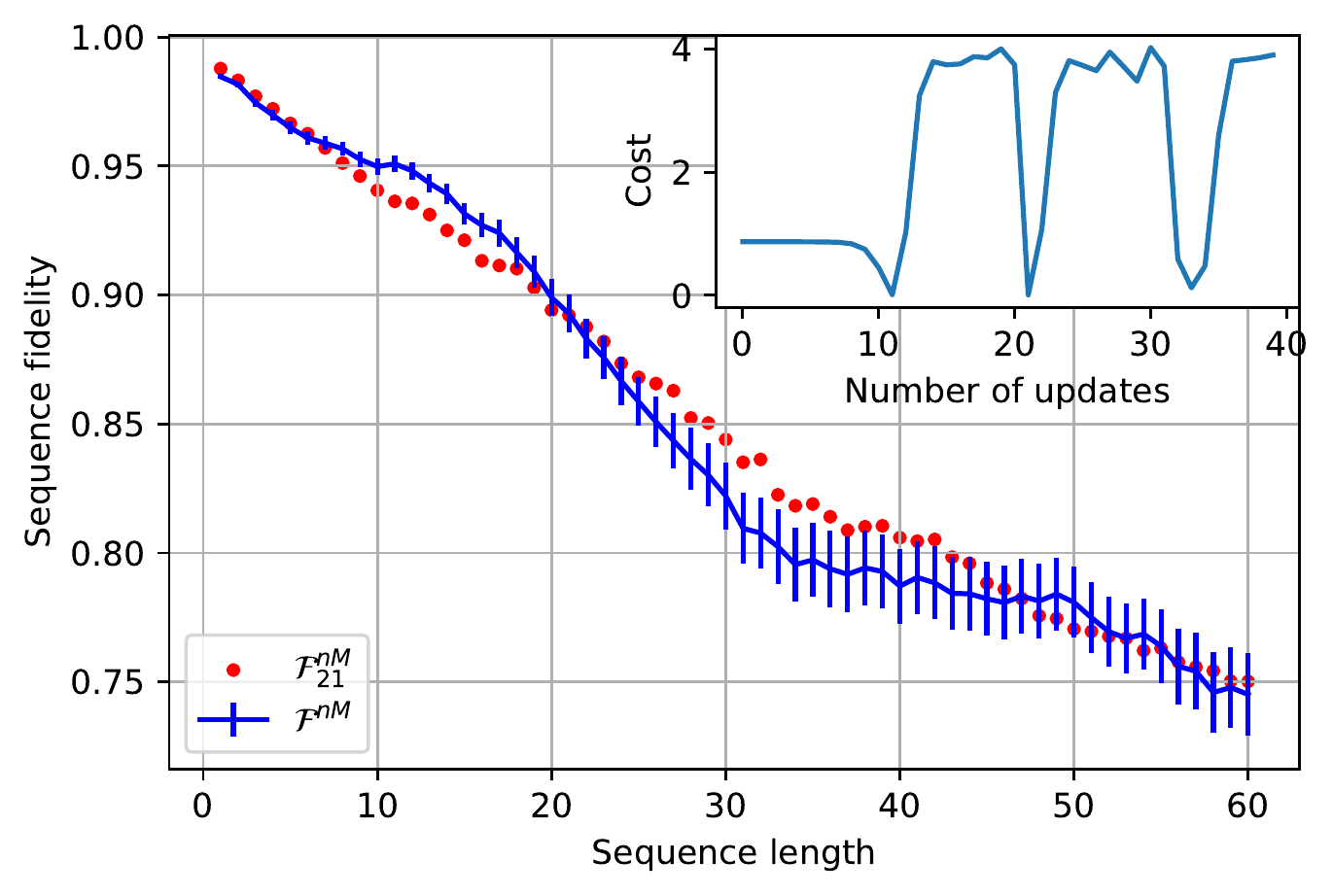}
    \caption{\textbf{Learned \gls{asf} for two qubit fully non-Markovain spin noise model with Adagrad.}
    The blue line is the \gls{asf} from the non-Markovian noise model and the error bar is the standard deviation divided by square root of sample numbers.
    The red dots are the output of our method, obtained by Adagrad with initail learning rate $0.001$.
    The superscript in $\mathcal{F}^{nM}_{21}$ indicates the noise model and the subscript is the number of iterations to reach this result.
    The subplot shows the change of cost with iterations, the $21^\textit{th}$ iteration gives the \gls{asf} data shown in here.}
    \label{fig:nM_asf_ada}
\end{minipage}%
\hfill
\begin{minipage}{.47\textwidth}
    \includegraphics[width=\linewidth]{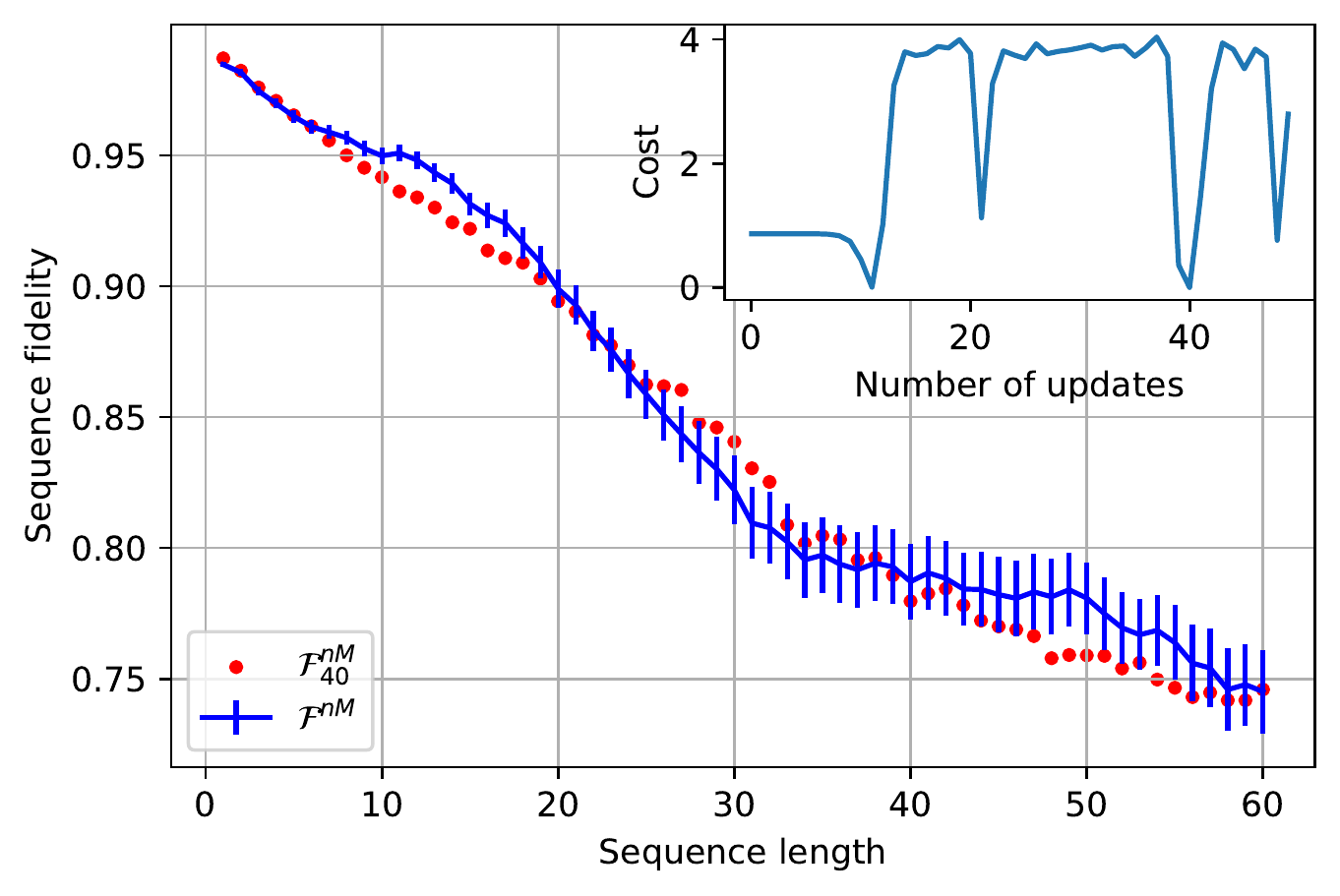}
    \caption{\textbf{Starting from the $21^{\textit{th}}$ update with different learning rate.}
    The blue line is the \gls{asf} from the non-Markovian noise model and the error bar is the standard deviation divided by square root of sample numbers.
    The red dots are the output of our method, starting with the initial noise from the $21^{\textit{th}}$ update in Fig.~\ref{fig:nM_asf_ada}. The result is obtained by Adagrad with initail learning rate $0.001001$.
    The superscript in $\mathcal{F}^{nM}_{40}$ indicates the noise model and the subscript is the number of iterations to reach this result.
    The subplot shows the change of cost with iterations, the $40^\textit{th}$ iteration gives the \gls{asf} data shown in here.}
    \label{fig:nM_asf_load_ada}
\end{minipage}
\end{figure}

We believe the sudden jump in the cost function might be caused by the way we impose the time-independent constraint (as we have discussed in the Discussion section in the main text, Section~\ref{sec: conclusions}, relaxing this comes with other challenges). The following two examples show that the jump is a common issue. Fig.~\ref{fig:nM_asf_ada} is obtained by Adagrad optimizer with initial learning rate $0.001$.
Since the cost fluctuates, we pick up the noise after the $21^{\textit{th}}$ update and start the optimization with a much small learning rate.
To our surprise, a small learning rate did not always fit the data better, it might update away from the minimum.
But with initial learning rate $0.001001$, a slightly bigger one, it is improved as shown in Fig.~\ref{fig:nM_asf_load_ada}.
This indicates that imposing the time-independent constraint while avoiding naively overfit the data and shortening the iterations, 
it makes the convergence unstable.

% \section{Toy examples}
% \subsection{Single-qubit Clifford gates with single environment qubit}
% \add{Apply to model of~\cite{figueroaromero2021randomised}}

% \subsection{Finite memory}
% \add{Toy model with correlated noise only up to a given time step}

% \subsection{Time dependent Markovian data}
% \add{What happens with Markovian time dependent decays?}

% \subsection{Non-Clifford gate sets}
% \add{Things should work regardless of the chosen group}

\end{document}